\documentclass[sigconf, table]{acmart}
\usepackage[marginal]{footmisc}

\AtBeginDocument{%
  \providecommand\BibTeX{{%
    \normalfont B\kern-0.5em{\scshape i\kern-0.25em b}\kern-0.8em\TeX}}}

\begin{document}
\begin{sloppypar}
\title{Emotion Recognition with Spatial and Functional Mapping of CNS and PNS Signals}

\author{Zhiyao Cen, Xiangwen Deng, Hengjie Zheng, Jianing Zhao, Anjie Jin, Chentao Fu, Tianqi Wang, Shangming Yang, Jingdian Yang}
 \affiliation{
   \institution{University of Electronic Science and Technology of China}}
\email{cenzhiyao, Xiangwen Deng, jianingzhao, anjiejin, fuchentao, tianqiwang@std.uestc.edu.cn}
\email{ minn003@163.com,      0015yang@gmail.com} 

\acmConference[Multimedia]{ACM Conference}{Oct 2022}{Lisbon, Portugal}


\begin{abstract}
  Emotion plays a significant role in our daily life. Recognition of emotion is wide-spread in the field of health care and human-computer interaction. Emotion is the result of the coordinated activities of cortical and subcortical neural processes, which correlate to specific physiological responses. However, the existing emotion recognition techniques failed to combine various physiological signals as one integrated feature representation. Meanwhile, many researchers ignored the problem of over-fitting model with high accuracy, which was actually false high accuracy caused by improper pre-processing. In this paper, sigmoid baseline filtering is conducted to solve the over-fitting problem from source. To construct a physiological-based algorithm, a 3D spatial and functional brain mapping is proposed based on human physiological mechanism and international electrode system, which combines the signals of the central and peripheral nervous system together. By combining the baseline filtering, 3D brain mapping, and simple 4D-CNN, a novel emotion recognition model is finally proposed. Experiment results demonstrate that the performance of the proposed model is comparable to the state of art algorithms.
\end{abstract}
  
  
  
  \maketitle

  \footnote{First Author and Second Author contribute equally to this work.\\}

  \section{Introduction}
  Positive emotion can improve our sense of happiness \cite{3}, while negative emotion may lead to health problems \cite{4}. Recently, emotion recognition has attracted more and more attention, especially in academic research.

  Generally, two main approaches are adopted for human emotion recognition: One utilizes non-physiological signals, such as facial expression \cite{5}, intonation \cite{6}, and emotional vocabulary \cite{7}. The another one utilizes physiological signals generated by the CNS and PNS (short for central and peripheral nervous system), including EEG \cite{11}, EOG \cite{14}, and PSR \cite{71}. As far as we are concerned, physiological signals are considered more reliable information, which reflect objective physiological variations of human emotion.

  However, many developed emotion recognition models focused on the accuracy but ignored the hidden problems in the pre-processing. Recently, a pre-processing method named base-mean, first proposed in a paper \cite{12}, was employed in many EEG-based emotional recognition research \cite{13,14,15,16,17,25,68}. Despite the fact that these models achieved high accuracy, it is further validated that this popular base-mean method results in a severe over-fitting problem which will finally lead to the unreliability of a model and false high accuracy. As validation, mathematical demonstration and experiment results are shown. To solve this over-fitting problem while following the idea of removing baseline effect, sigmoid baseline filtering in frequency domain is proposed.

  On the other hand, there has never been a reliable and physiological-based algorithm to combine different physiological signals as an integrated representation. Some researchers only adopted EEG for emotion recognition \cite{71,73}, some roughly combined EEG with other physiological signals after separate feature extraction \cite{14,62}. To construct a highly reliable integrated feature representation, a 3D spatial and functional brain mapping algorithm is proposed in this paper, following the idea of cortical cartography \cite{83}. In this mapping algorithm, EEG signal is mapped as CNS signal by following the electrode system \cite{82}, and the selected peripheral physiological signals are mapped as PNS signals by coordinating the research results of functional brain divisions \cite{42,45,46,47,81}. Finally, a simple 4D-CNN is applied to extract features of CNS and PNS signals simultaneously on spatial, temporal, and functional dimensions.

  In summary, the main contributions of this paper are as follows:

  (1) The over-fitting problem of the popular base-mean is revealed, and sigmoid baseline filtering is proposed to solve the problem.

  (2) The CNS and PNS signals are integrated into one 3D representation for the first time, by applying the proposed spatial and functional brain mapping based on physiological research results and international electrode system.

  (3) Extensive comparative experiments on DEAP and SEED datasets are conducted to demonstrate that the proposed model has comparable or better performance in emotion recognition.

  \begin{figure*}[htb]
  \includegraphics[width=\textwidth]{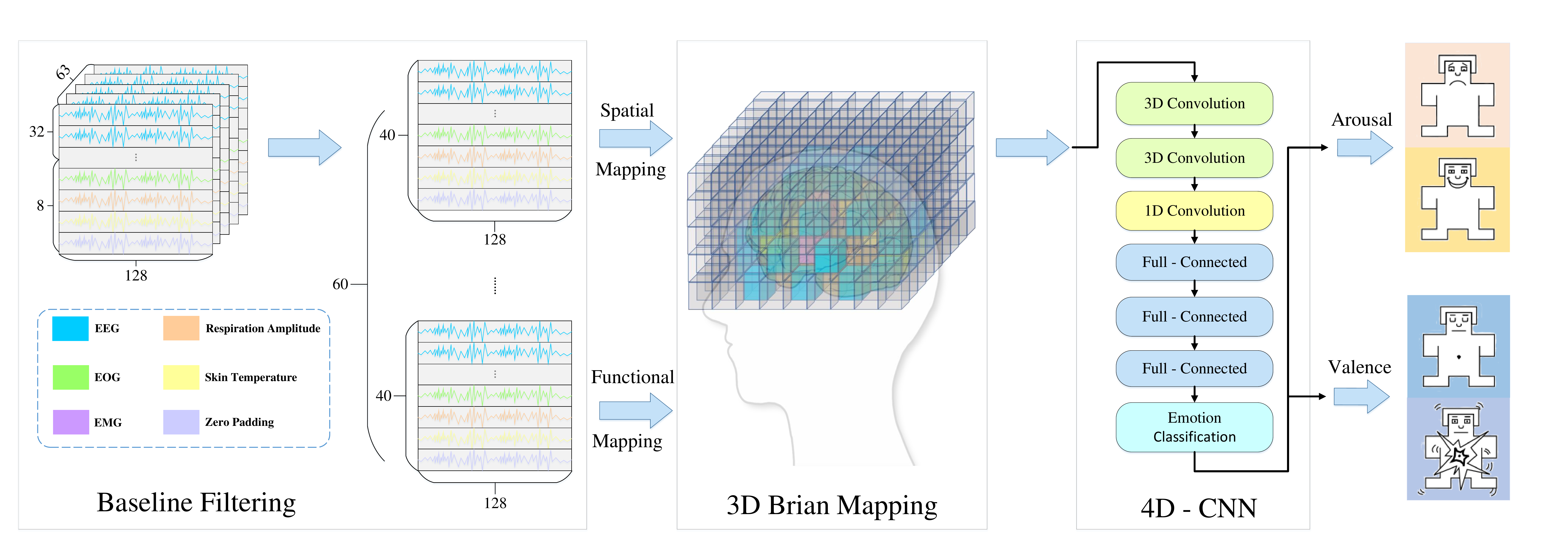}
  \caption{The overview framework of the proposed method. Firstly, the sigmoid baseline filtering is proposed as substitute for an over-fitting base-mean algorithm. Secondly, the spatial and functional brain mapping of CNS and PNS signals is employed as an integrated feature representation. Finally, a simple 4D-CNN is applied for spatial, temporal, and functional feature extraction and emotion classification.}
  \end{figure*}
  
  \section{Related Works}
  The literature related to our model can be roughly divided into two categories.
  
  \subsection{Pre-processing of CNS and PNS signals}
  
  Pre-processing is absolutely necessary for neuron network models. Both CNS and PNS signals are discrete timing signals, which record electrical intensity varying with time.
  
  In \cite{25}, Z-sore normalization was implemented to normalize values per frame. In \cite{14}, different temporal window sizes were compared for EEG and EOG signals. Different band pass filtering in frequency domain were applied in \cite{12,55,78}. Besides that, one pre-processing with baseline called base-mean, proposed by Yang in \cite{12}, stood out for its high accuracy. This base-mean method tends to remove baseline effect in raw EEG signals. But it also unintentionally results in a high similarity problem, an over-fitting model, and the false high accuracy result. To overcome the above over-fitting problem while following the idea of baseline effect removal, the sigmoid baseline filtering in frequency domain is proposed.

  \subsection{Feature Representation and Extraction of CNS and PNS signals}
  
  The emotional features in CNS and PNS signals can be extracted on three dimensions: spatial, temporal, and functional. Many researchers focused on spatial or temporal dimension and proposed various methods of feature representation and extraction, but very few of them made the best of PNS signals on the functional dimension.

  On temporal dimension, CNS and PNS signals are time-series data in essence. So, many models were aimed at constructing and extracting temporal features. ACRNN in \cite{67} was based on RNN and self-attention mechanism. Scaling-Net in \cite{70} deployed convolution layers to extract temporal features. FAWT-RF in \cite{73} transformed data to frequency domain for noise reduction and emotion classification.

  On spatial dimension, emotional features also exist in the complex correlations between different EEG channels. To reconstruct the relative positional relationships of EEG signals, the form of 2D image series was put forward. Then, different interpolation algorithms were applied to further reconstruct the missing information \cite{26,27}. Meanwhile, high-level spatial features were also revealed: Symmetric features were extracted by combining different ways of folding \cite{16}, and deeper relationships between adjacent channels were revealed by using dynamic and graphic EEG representation \cite{28}.

  On functional dimension, the multi-modal data can be adopted as auxiliary signals with EEG in emotion recognition task. Multi-modal data or peripheral physiological signals include EOG, EMG, GSR and so on\cite{11}, which reflect objective variations of human physiological function. However, existing models with multi-modal data failed to take CNS and PNS signals as an integrated feature representation \cite{62,75}. In our proposed model, CNS signals and PNS signals are merged in 3D cuboid series using spatial and functional mapping, which maximizes the potential of all physiological signals.

\section{Proposed Method}
  
  The proposed model is composed of three parts: the sigmoid baseline filtering, the spatial brain mapping of CNS signals, the functional brain mapping of PNS signals, and a simple 4D-CNN. The overview framework of our proposed model is shown in Fig 1.

  \subsection{Pre-processing with Baseline Signals}
  
  Pre-processing is essential for neuron networks, and baseline-related pre-processing of EEG signals was already conducted in research field \cite{14}. Recently, one popular pre-processing called base-mean, which was first proposed by Yang \cite{12}, was employed in many EEG-based emotion recognition research \cite{13,15,16}. However, it is later proved that this base-mean method leads to severe over-fitting problem.

  \subsubsection{Strengths and Weaknesses of Base-mean Pre-processing}
  \
  \newline
  \indent Base-mean pre-processing is a simple and effective method to remove the baseline effect or the resting state in EEG signals. But it also results in extra similarity between training set and test set, which leads to an over-fitting model with false high accuracy result.
  
  The strengths of base-mean are its simplicity and effectiveness of removing baseline effect in raw EEG data. As to simplicity, the base-mean only contains simple shape transformation and common mathematical operation. As to effectiveness, many models achieved very high accuracy by using base-mean method: The accuracy of LSTM-CNN in \cite{12} improved greatly to 97.8\%, 91.03\% for arousal and valence classification. Additionally, other models using base-mean method also achieved very high accuracy: 96.61\%, 96.43\% in 3D-CNN \cite{25}, and 94.22\%, 94.58\% in 4D-Attention \cite{68}).
  
  
  The major weaknesses of base-mean method are the similarity and over-fitting problems. Crucial steps of base-mean are explained below. In the beginning, the Base-Mean and Raw-EEG matrices of the same shape are the results of splitting and taking element-wise mean values of the EEG data. Then Base-Removed matrices are equal to subtraction of Raw-EEG with Base-Mean. On the surface, the subtraction is considered as removal of baseline effect. But actually, it is also a way of baseline marking, as the homologous Raw-EEG matrices, which correspond to the same emotion label, are marked by the same Base-Mean matrix. Thus, similarity between homologous Base-Removed matrices increases for their same baseline feature. Then, the random shuffling separates homologous Base-Removed matrices into a training set and test set randomly. Finally, the extra baseline marking can be learned easily by the neuron network, which leads to over-fitting and false high accuracy.

  \begin{table*}[htb]
  \centering
  \caption{The similarity between matrices after pre-processing}
  \begin{tabular}{lcccc}
  \toprule
  \multicolumn{1}{l}{Matrices} & \multicolumn{1}{l}{Euclidean Distance(\%)} & \multicolumn{1}{l}{Cosine Similarity(\%)} & \multicolumn{1}{l}{Pearson Coefficient(\%)} \\
  \midrule
  within Raw-EEG & $1.12$±$1.04$ & $0.29$±$0.30$ & $0.64$±$0.41$ \\
  Base-Mean and Raw-EEG & $1.05$±$0.84$ & $0.01$±$0.16$ & $0.50$±$0.51$ \\
  &  &  &  &  \\
  within Base-Removed & $1.12$±$1.04$ & $0.46$±$0.28$ & $0.73$±$0.34$ \\
  Raw-EEG and Base-Removed & $0.47$±$0.43$ & $0.86$±$0.13$ & $0.93$±$0.12$ \\
  Base-Mean and Base-Removed & $0.91$±$0.77$ & $0.46$±$0.22$ & $0.73$±$0.31$ \\
  &  &  &  &  \\
  within Filtered-EEG & $1.22$±$1.14$ & $0.34$±$0.32$ & $0.67$±$0.40$ \\
  Raw-EEG and Filtered-EEG & $0.45$±$0.44$ & $0.86$±$0.13$ & $0.93$±$0.11$ \\
  Base-Mean and Filtered-EEG& $1.04$±$0.89$ & $0.16$±$0.33$ & $0.58$±$0.48$ \\
  \bottomrule
  \end{tabular}
  \end{table*}

  In Table 2, the results validate the existence of the high similarity problem and the baseline marking. On the one hand, base-mean indeed removes the baseline effect, for the similarity within homologous Base-Removed matrices actually increased compared to Raw-EEG. On the other hand, the notable increase of similarity between Base-Removed and Base-Mean validates the existence of baseline marking.

  Matrix similarity indexes include Euclidean Distance $(S_1)$, Cosine Similarity $(S_2)$, and Pearson Correlation Coefficient $(S_3)$, which are calculated as follows,

  \begin{eqnarray}
  S_{1} = \sqrt{\sum_{i = 1}^{m} \sum_{j = 1}^{n}\left | A_{ij}-B_{ij}  \right | ^{2} }
  \end{eqnarray}

  \begin{eqnarray}
  S_{2} = \frac{\sum_{i = 1}^{m} \sum_{j = 1}^{n}\left ( A_{ij} \times B_{ij}  \right ) }{\sqrt{\sum_{i = 1}^{m} \sum_{j = 1}^{n}\left ( A_{ij}  \right )^{2}\times \sum_{i = 1}^{m} \sum_{j = 1}^{n}\left ( B_{ij}  \right )^{2}  }}
  \end{eqnarray}

  \begin{eqnarray}
  S_{3} = \frac{COV\left ( A, B \right ) }{(\sigma A) \times(\sigma B)},
  \end{eqnarray}
  where A and B represents two matrices of the same shape, m and n represents number of channels and frames per data, cov represents the covariance, and $\sigma$ represents standard deviation.

  In Table 1, the results of DT, KNN, and SVM are listed, showing the severe over-fitting problem and false high accuracy by using base-mean method. The 1st and 2nd rows show that the accuracy is very high even with minimal training ratio by using base-mean, which is usually a sign of over-fitting. The 2nd, 3rd, and 4th rows show that the model using base-mean actually learned a lot from the baseline features in the homologous Base-Removed matrices. The 5th and 6th rows show that the severe over-fitting results by using base-mean in the way of replacing EEG data with random-valued but normalized data. All these results prove that base-mean pre-processing leads to over-fitting and false high accuracy.

  \begin{table}
  \centering
  \caption{Model results using base-mean pre-processing}
  \begin{tabular}{lcccc}
  \toprule
  Method & Split (ratio) & Arousal(\%) & Valence(\%) \\
   & & DT/KNN/SVM & DT/KNN/SVM \\
  \midrule
  Raw & random(0.8) & $0.74$/$0.78$/$0.76$ & $0.78$/$0.76$/$0.72$ \\
  Base-mean & random (0.2) & $0.78$/$0.97$/$0.73$ & $0.96$/$0.97$/$0.95$ \\
  &  &  &  &  \\
  Base-mean & by-data (0.8) & $0.49$/$0.59$/$0.56$ & $0.61$/$0.57$/$0.58$ \\
  Base-mean & by-index (0.2) & $0.77$/$0.97$/$0.69$ & $0.94$/$0.98$/$0.96$ \\
  &  &  &  &  \\
  Random   & by-data (0.8) & $0.49$/$0.35$/$0.88$ & $0.62$/$0.65$/$0.67$ \\
  Random   & by-index (0.2) & $0.99$/$1.00$/$0.99$ & $1.00$/$1.00$/$1.00$ \\
  \bottomrule
  \end{tabular}
  \end{table}

  In the table, Raw represents EEG data without base-mean pre-processing, Random represents usage of random-valued but normalized matrices, and Split (ratio) represents ways of train-test-split and the training set ratio. Splitting by data means dividing the homologous matrices generated from one original data all to the training set or all to the test set, whereas the splitting by index means randomly dividing the homologous matrices from one original data between the training and test set based on the exact ratio.

  \subsubsection{Sigmoid Baseline Filtering}
  \
  \newline
  \indent The over-fitting problem of base-mean stems from dealing with different homologous Raw-EEG matrices by subtracting them from the same Base-Mean matrix. To solve the high similarity and over-fitting problem from source, the sigmoid baseline filtering is conducted in frequency domain, following the idea of baseline effect removal, multiple band-pass filter, and echo reduction in multiple microphone audio system. \cite{35,36}. The filtering mainly contains two steps: First, convert data to frequency domain and quantify the baseline effect; Second, remove the baseline effect and convert data back to time domain.

  The raw data are divided into Base-Mean and Raw-EEG matrices, which are transformed into frequency domain by using Fast Fourier Transform (FFT) as follows,

  \begin{eqnarray}
  RT,BT=FFT(R), FFT(B),
  \end{eqnarray}
  
  \noindent where R and B represents the Raw-EEG and Base-Mean matrices, while RT and BT represents Raw-EEG and Base-Mean matrices after FFT.
  
  Then, the baseline influence can be measured by the difference between the values of RT and BT, forming the Deactivate-Filter matrices which are calculated as follows,
  \begin{eqnarray}
  D=Sig(BT-RT),
  \end{eqnarray}
  
  \noindent where D represents Deactivate-Filter matrix containing quantified value of baseline effect in every Raw-EEG matrix. Sig means the sigmoid function which softens the result. Notice that Deactivate-Filter is computed differently even for homologous RT. So the high similarity problem is avoided, and the over-fitting problem is solved from source.

  Finally, the Filtered-EEG matrices are calculated using above RT, BT, and D matrices and transformed back to time domain, which is shown as follows,

  \begin{eqnarray}
  F=IFFT(RT-D\circ BT),
  \end{eqnarray}
  \noindent where F represents Filtered-EEG, symbol $\circ$ indicates element-wise multiplication (Hadamard product). IFFT means Inverse Fast Fourier Transform (IFFT).

  To summarize, the core idea of sigmoid baseline filtering is to remove baseline effect from every segment of raw EEG. Specifically, to lower the value of some frequencies in Raw-EEG, which have high intensity in Base-Mean. Unlike base-mean, the over-fitting problem is avoided from source for every homologous EEG segments are dealt with differently. The similarity comparison between matrices using sigmoid baseline filtering is shown in Table 2.

  \subsection{Spatial and Functional Brain Mapping of CNS and PNS signals}
  
  To construct a reliable and integrated feature representation for both CNS and PNS signals, the 3D spatial and functional brain mapping algorithm is proposed, following the concept of the cortical cartography \cite{83}. This brain mapping algorithm has three steps: First, map the CNS signals (EEG) into cuboid series; Second, select appropriate peripheral signals as PNS signals; Third, map the selected PNS signals into the 3D series along with CNS signals. This mapping algorithm is based on the literature of functional brain division \cite{42,45,46,47,81} and 10-20 international electrode placement system \cite{82}.

  \begin{figure*}[htb]
    \centering
    \includegraphics[width = 0.95 \linewidth]{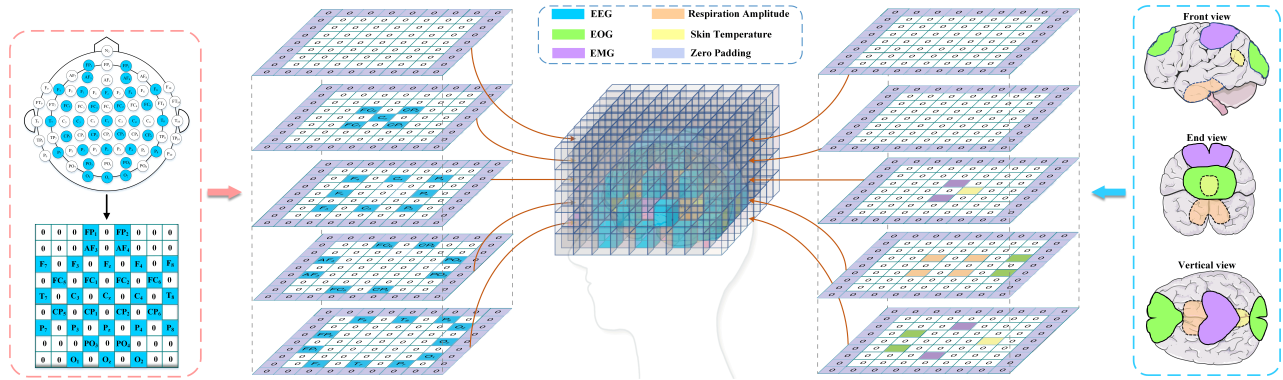}
    \caption{The spatial and functional brain mapping. The left side is the spatial mapping of CNS signals (EEG), where names like $FC_1$ represents EEG electrode channels. The right side is the functional mapping of PNS signals, which includes signals of EOG, EMG, Respiration Amplitude, and Skin Temperature. Mapping locations of PNS and CNS signals are referencing 10-20 international electrode placement system and brain division literature.}
    \end{figure*}

  \subsubsection{Spatial Brain Mapping of CNS signals}
  \
  \newline
  \indent Spatial mapping of EEG signals to 2D EEG image series was employed extensively in many studies, which reconstructed the EEG representation. Based on the 2D image-like representation of EEG, different interpolation algorithms were employed to obtain the missing information of adjacent channels. From this perspective, the 3D spatial mapping of CNS signals is similar to the 2D image representation, further simulating the spatial structure of EEG.
  
  However, the purpose of the proposed spatial mapping is to reconstruct the upper cerebral hemisphere with CNS signals while sparing empty space for the functional mapping of PNS signals. The mapping locations of CNS signals in the cuboid are considered as anchor points, which determine the center and boundary of functional brain divisions. Thus, the mapping locations of PNS signals can be calculated. The proposed 3D spatial mapping is shown on the left side of Fig 2.
  
  In the figure, names like $FC_1$ represent EEG electrode channels, and characters like F, Fp, T, C, O, and P represent frontal, prefrontal, temporal, central, occipital, and parietal lobes. The odd and even number subscript of characters respectively represent left and right hemispheres of brain.
  
  \subsubsection{Selection of PNS signals}
  \
  \newline
  \indent In our proposed model, the essence of PNS functional brain mapping is to select appropriate physiological signals as PNS signals, and map them into the 3D cuboid series. As for PNS signals selection, electrode placement system and functional brain division literature are referenced.

  On DEAP dataset, there are 8 types of PNS signals: GSR, Respiration amplitude, Skin temperature, Blood volume by plethysmograph, Electromyogram (EMG) of Zygomaticus and Trapezius muscles, and Electrooculogram (EOG) of vertical and horizontal. Peripheral signals with apparent brain division and close to emotion are preferentially chosen to be the final PNS signals.
  
  EMG is an auxiliary examination of some neurological and muscular diseases \cite{40,41}. It reflects the health and excitability of nervous system related to muscle movement, which relates to motor cortex located in anterior central gyrus at the boundary of frontal and central lobes \cite{42}.
  
  Respiration amplitude was adopted in many emotion-related studies, like feedback of Laughter \cite{43} and sleep \cite{44}. The respiration center \cite{81}, which gives orders to respiratory muscles, is located in the medulla oblongata and pons, which is close to the bottom of brain.
  
  \begin{table}
    \centering
    \caption{The brain regions and electrodes of PNS signals}
    \begin{tabular}{ccc}
    \toprule
    Type & Lobes & Electrodes \\
    \midrule
    EOG(horizontal) & Frontal & $FP_1$,$F_3$,$F_z$,$AF_3$ \\
    & Occipital,Pariental & $PO_3$,$O_1$,$O_z$ \\
    EOG(vertical) & Frontal & $FP_2$,$F_4$,$F_z$,$AF_4$ \\
    & Occipital,Pariental & $PO_4$,$O_2$,$O_z$ \\
    EMP(zygomaticus) & Central(left) & $FC_1$,$FC_5$,$CP_1$,$CP_5$ \\
    & Central(right) & $FC_2$,$FC_6$,$CP_2$,$CP_6$ \\
    EMG(trapezius) & Central(left) & $FC_1$,$CP_1$,$C_z$ \\
    & Central(right) & $FC_2$,$CP_2$,$C_z$ \\
    Skin temperature & Central(Occipital) & $CP_1$,$PO_3$,$CP_2$,$PO_4$ \\
    Respiration amplitude & Central(Bottom) & (Near Brain Bottom) \\
    \bottomrule
    \end{tabular}
    \end{table}
  
  Skin temperature was also adopted in many emotional recognition tasks \cite{45}. The preoptic anterior hypothalamus (POAH) is the principal center for body temperature regulation \cite{46}, which is close to the center of brain.
  
  EOG records electrical signals of vertical and horizontal eye movements, which was essential in some emotion recognition models \cite{75}. Since many brain regions are involved with eye movement and human vision, two visual cortex in the frontal lobe and pariental lobe are both considered \cite{47}.
  
  GSR is short for Galvanic Skin Response. It is usually regarded as a measurement of perspiration, which reflects stress and surprise \cite{48}. Studies showed that GSR is related to arousal \cite{49}, but no corresponding brain region is found for GSR.
  
  Electrocardiogram was also adopted in the field of emotion recognition. But the existence of cardiac autonomic nervous means that heart functions quite independently, and there is no brain region specifically associated with it.
  
  Blood volume is an essential index in medicine. However, it is influenced by many factors in complex ways, so no specific brain region is related. Besides that, the correlation between blood volume and emotion is also not yet verified.
  
  As mentioned above, EOG, EMG, Respiration amplitude, and Skin temperature are selected as PNS signals for further functional mapping.

  \subsubsection{Functional Brain Mapping of PNS signals}
  \
  \newline
  \indent After the selection of peripheral physiological signals, the functional brain divisions of different PNS signals are also evident. Then, the mapping locations for PNS signals can be calculated, referencing the 10-20 international electrode placement system and functional brain division literature, while using mapped CNS channels as anchor points. The proposed 3D functional mapping is shown on the right side of Fig 2.

  First, the brain regions are evident for PNS signals according to the physiological research results related to peripheral signals. This process is elaborated in 3.2.2 and recorded as follows,
  
  \begin{eqnarray}
  BR=GetRegion(T),
  \end{eqnarray}

  \noindent where GetRegion function returns the brain region (BR) of a given type (T) of PNS signals.
  
  Second, find the standard electrodes in the corresponding brain region. The center of brain region is calculated as follows,
  \begin{eqnarray}
  DPC=Avg(GetElectrodes(BR)),
  \end{eqnarray}

  \noindent where GetElectrodes function returns the electrode locations according to the given brain region, DPC represents the center of the brain region, and Avg function returns integer-rounded center location of given electrode locations while avoiding overlapping with the mapped CNS signals. Table 3 shows the corresponding brain divisions and standard electrodes of PNS signals, where names like $FC_1$ represent name of EEG electrode channels, according to the 10-20 international electrode placement system.
  
  Third, the final mapping locations of PNS signals are calculated as follows,

  \begin{eqnarray}
  MP=Avg(DPC,CP),
  \end{eqnarray}

  \noindent where CP represents the brain center which is close to the hypothalamus, and MP represents the final mapping locations of PNS signals. The reason for using brain center is that the CNS signals are recorded on the brain surface, so the mapping of PNS signals should be closer to the brain center.

  \subsubsection{4D-CNN based Feature Extraction}
  \
  \newline
  \indent Many researchers employed neuron networks to recognize emotion because of the outstanding performance of deep learning approaches in pattern recognition tasks. Among all these advanced deep learning methods (GCNN, LSTM, GCN, SNN, and Caps-Net), convolution neuron network is considered most appropriate, for the proposed model contains 4D features lying in spatially adjacent channels and temporally continuous frames. Since the convolution kernels have multiple dimensions, emotional features in the 4D data can be easily extracted.

  After the spatial and functional mapping of CNS and PNS signals, the data is in the form of 3D series. Then, a relatively simple and orderly 4D-CNN is proposed, mainly composed of three kinds of layers: two 3D-Conv layers for spatial feature extraction, one 1D-Conv layer for temporal feature extraction, and the final three fully connected layers for emotion classification. Meanwhile, RELU, batch-norm, and drop out layers are also employed to improve training efficiency and model stability. The implementation details are in chapter 4.2.
  
  \section{Experiments}
  
  \subsection{Materials}
  
  The DEAP dataset is selected to evaluate the performance of proposed model, which is extensively used in many EEG-related studies. Comparative experiments on SEED dataset are also conducted, which is shown in Chapter 5.3.3.
  
  The DEAP \cite{11} is an open dataset developed by Koelstra and colleagues. This dataset contains 32 EEG channels and 8 peripheral physiological channels, which were collected when 32 participants watched 40 videos, each with 60 seconds duration. Each data contains 63s signals, and the first 3s are the baseline signals. The participants rated a self-assessment of arousal, valence, liking, and dominance after watching 60s video. The signals are down-sampled to 128hz, and the data size of DEAP is (32 $\times$ 40 $\times$ (32+8) $\times$ (128 $\times$ 63)), which represents (participants $\times$ videos $\times$ channels $\times$ frames). The structure of the DEAP dataset is shown in Table 4.

  \begin{table}[htb]
  \centering
  \caption{The structure of DEAP dataset}
  \begin{tabular}{ccc}
  \toprule
  Array name & Array shape & Array concents \\
  \midrule
  Data & $40\times40\times8064$ & $Videos\times Channels\times Frames$\\
  Labels & $40\times4$ & $Videos\times Labels$ \\
  \bottomrule
  \end{tabular}
  \end{table}

  \subsection{Implementation Details}
  
  For pre-processing, data is processed with some basic algorithms before the sigmoid baseline filtering. Z-score algorithm is used to normalize every temporal frame, making it easier for network to train. Dualization of emotion labels is utilized for two-class classification task on DEAP dataset (score in range [5, 9) is positive and [1, 5) is negative).

  \begin{figure}[htb]
  \centering
  \includegraphics[width = 0.95 \linewidth]{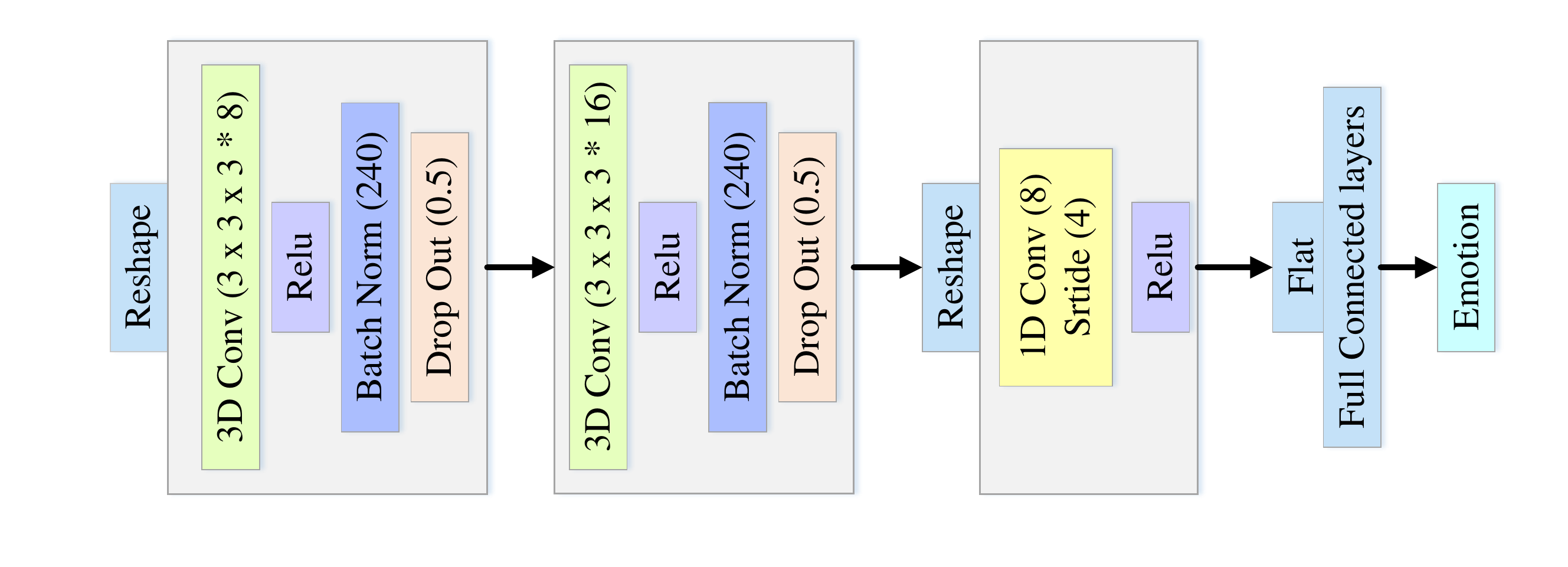}
  \caption{The structure of 4D-CNN}
  \end{figure}
  
  \begin{table*}[]
  \centering
  \caption{The comparison between models in subject-independent emotion recognition}
  \begin{tabular}{cccccc}
  \toprule
  Referrence & Model Name & DEAP Arousal(\%) & DEAP Valence(\%) & Base-mean & Year \\
  \midrule
  \cite{71} & 2D-CNN &          76.56 & 80.46 & N & 2018  \\
  \cite{12} & CNN-LSTM &        91.03 & 90.80 & Y & 2018  \\
  \cite{73} & FAWT-RF &         79.95 & 79.99 & N & 2019  \\
  \cite{74} & MC-CNN &          88.49 & 87.44 & N & 2019  \\
  \cite{77} & CNN-SAE-DNN &     89.49 & 92.86 & N & 2020  \\
  \cite{25} & 3D-CNN &          96.61 & 96.43 & Y & 2020  \\
  \cite{68} & 4D-Attention &          94.22 & 94.58 & Y & 2021  \\
  \cite{17} & SFE-Net &         91.94 & 92.49 & Y & 2021  \\
  \cite{14} & EEG-EOG &         86.38 & 85.42 & N & 2021  \\
    & \textbf{Ours} &  \textbf{92.36} & \textbf{92.64} & \textbf{N} & \textbf{2022}  \\
  \bottomrule
  \end{tabular}
  \end{table*}

  For the 4D-CNN network, details of parameters and layers are listed below. To extract spatial and temporal features in 3D series data (frames $\times$ (length $\times$ width $\times$ height)), the first two layers are 3D convolution layer, which is followed by one 1D convolution layer, and three fully connected layers are adopted for the final emotion classification. Traditionally, RELU is considered as an excitation function. Dropout layer with 0.5 dropping rate and batch-normalization layer are implemented after each 3D convolution layer. Zero padding has been already implemented in the stage of 3D brain mapping. The kernel size is (1 $\times$ (3 $\times$ 3 $\times$ 3)) in 3D convolution and (8 $\times$ (1 $\times$ 1 $\times$ 1)) in 1D convolution. The stride in 1D convolution is (4,1,1,1). The number of feature maps are 8 and 16 for two 3D convolution layers, which can simplify and accelerate the training process. The Adam optimizer and the cross-entropy loss function are chosen to calculate and minimize the loss. The learning rate is set to 0.001 and L2 penalty parameter is set to 0.001. The training epoch number is set to 100 and batch number is set to 240. Five-fold cross validation is considered in the experiments. The structure of 4D-CNN is shown in Fig 3.

  \section{Results and Discussion}
  
  \subsection{Results for Subject-dependent Classification}
  
  To further verify the effectiveness of the proposed model, the subject-independent comparative experiments are conducted between our model and the latest models, including hybrid neuron network, multi-modal network, attention network and so on.
  
  Both CNN-SAE-DNN \cite{77} and CNN-LSTM \cite{12} are hybrid network models, which can be used to extract features on multiple dimensions. EEG-EOG \cite{14} combines EEG and EOG signals for emotion recognition. 2D-CNN \cite{71} is implemented to extract features of EEG and GSR on spatial and temporal dimensions. 4D-CNN \cite{68} is based on 4D feature representations, and the attention mechanism is applied. SFE-Net \cite{16} applied different folding algorithms to extract symmetric EEG features. To balance between fairness and effectiveness in comparison, all the accuracy results of models are from their own literature.
  
  The accuracy comparison of arousal and valance classification on DEAP dataset are shown in Fig 4 and Fig 5, from which we can see that: First, the proposed model is superior to many CNN-based networks (MC-CNN and CNN-SAE-DNN) and a traditional machine learning model (FAWT-RF); Second, our model performs the best among other models using multi-modal data (2D-CNN and EEG-EOG); Third, the proposed model is still competitive with models applying base-mean pre-processing with false high accuracy (CNN-LSTM, SFE-Net). The specific five-fold results of proposed model for emotion classification on DEAP dataset are 92.22\%, 92.31\%, 92.17\%, 92.13\%, and 93.18\% for arousal and 92.46\%, 92.68\%, 92.33\%, 92.68\%, and 93.06\% for valence.

  \begin{figure}
    \centering
    \includegraphics[width = 0.95 \linewidth]{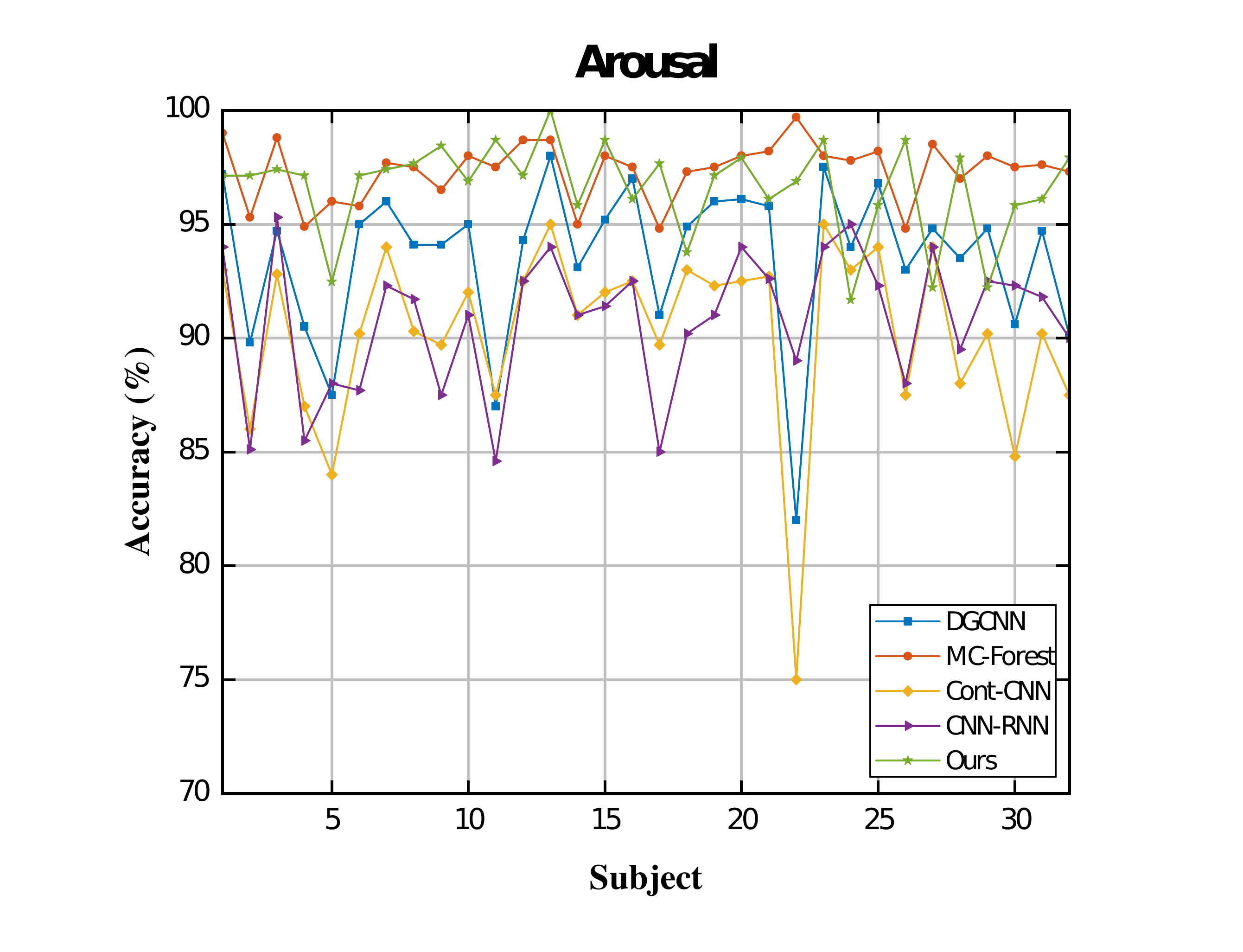}
    \caption{The arousal classification accuracies (\%) of all subjects}
    \end{figure}
    
    \begin{figure}
    \centering
    \includegraphics[width = 0.95 \linewidth]{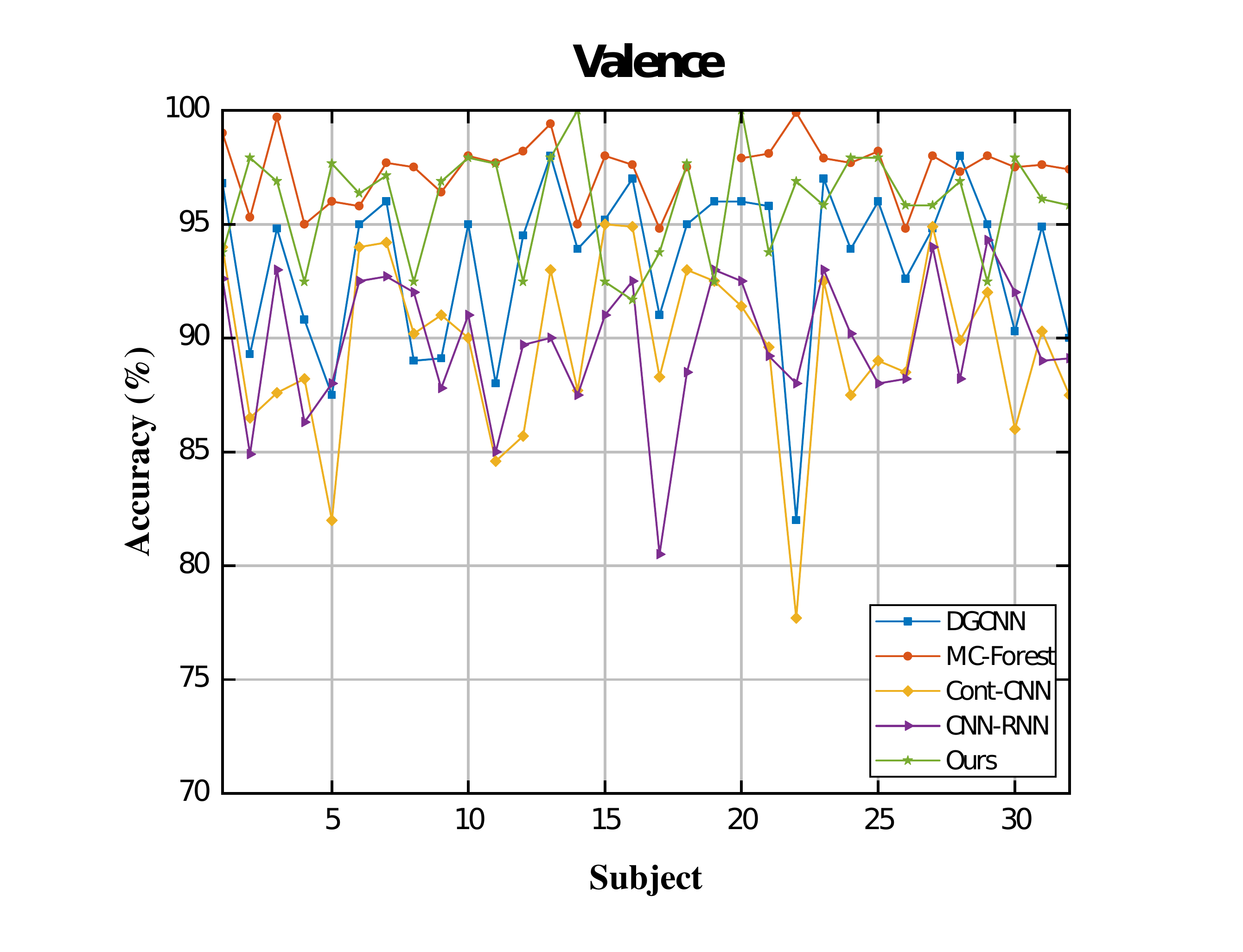}
    \caption{The valence classification accuracies (\%) of all subjects}
    \end{figure}

  \subsection{Results for Subject-independent Classification}
  
  To show the comparable performance of the proposed model, testing results of proposed model and state-of-the-art models are compared. Models include dynamic graph convolutional neuron network (DGCNN) \cite{61}, multi-grain cascade forest (MC-Forrest) \cite{79}, continuous CNN (Cont-CNN) \cite{53} and CNN-RNN \cite{81}. Generally, the proposed model shows obvious advantages in subject-dependent emotion classification: The mean value of 32 participants' classification accuracy is 96.62\% for arousal and 95.89\% for valence.
  
  In DGCNN, the dynamic graphic convolution helps to iterate and find the correlation between EEG channels. In MC-Forest, an enhanced deep random forest is employed to extract complex features. In Cont-CNN, baseline signals are utilized, and continuous convolution layers are implemented. In CNN-RNN, both spatial and temporal characteristics are extracted. Our proposed model benefits from the baseline and constructs abundant spatial functional features. The subject-dependent experiment results are shown in Table 5.
  
  From the table, the accuracy of the proposed model is about 10\% higher than Cont-CNN and CNN-RNN, more stable than DGCNN, and roughly the same as MC-Forest. The results show that our model with baseline filter and brain mapping is very efficient and effective for individual emotion recognition.

  \subsection{Discussion}
  
  In this section, three groups of comparative experiments are conducted to further evaluate the performance of the proposed method. The experiments include: Compare different levels of spatial and functional brain mapping to prove the effectiveness; Compare different combinations of 3D and 1D convolution layers to find the best fit; Compare different levels of 3D spatial mapping on SEED dataset to fully explore the possibility of 3D brain mapping. For all comparative experiments, the training epoch is set to 50, while other settings, processes, and structure remain unchanged as much as possible.

  \subsubsection{Different levels of spatial and functional brain mapping}
  \
  \newline
  \indent To show the advantages of spatial and functional brain mapping of CNS and PNS signals, four levels of brain mapping are compared: 2D image EEG series, 3D series after spatial brain mapping, 3D series after spatial and functional brain mapping (the proposed model), 3D series lacking of specific PNS signal.

  \begin{table}[htb]
  \centering
  \caption{Different levels of brain mapping}
  \begin{tabular}{lcc}
  \toprule
  Structure &        Arousal(\%)  &  Valence(\%)  \\
  \midrule
  2D image        &  $83.61$±$0.05$ &  $84.86$±$0.05 $ \\
  3D cuboid      &  $75.01$±$0.06$ &  $76.79$±$0.06 $ \\
  Our model       &  $91.95$±$0.04$ &  $91.11$±$0.04 $ \\
   & &\\
  Without EOG     &  $86.82$±$0.05$ &  $85.91$±$0.06 $ \\
  Without EMG     &  $85.29$±$0.05$ &  $84.70$±$0.04 $\\
  Without Resp    &  $91.47$±$0.04$ &  $91.69$±$0.03 $ \\
  Without Temp    &  $89.62$±$0.04$ &  $89.98$±$0.04  $\\
  \bottomrule
  \end{tabular}
  \end{table}
  
  The comparison results in Table 6 show that the 3D spatial and functional mapping is indeed effective, achieving about 7\% higher accuracy than 2D image EEG series. However, the results also indicate that the spatial mapping alone is less efficient than 2D series. This phenomenon is actually results from the inadequate number of EEG channels on DEAP dataset (32 in total), which is further validated in chapter 3.2 on SEED dataset. The results also show that EOG and EMG are the most effective PNS signals, while the respiration amplitude seems to be the least useful.
  
  \subsubsection{Different combinations of convolution layers}
  \
  \newline
  \indent To show the effectiveness of the developed 3D plus 1D convolution layers and find its best fit, three different combinations of 3D and 1D convolution layers are compared: one 3D convolution layer, two 3D convolution layers, one 3D with one 1D convolution layers, and two 3D with one 1D convolution layers.

  \begin{table}[htb]
  \centering
  \caption{Various combinations of convolution layers}
  \begin{tabular}{lcccc}
  \toprule
  Methods & \multicolumn{1}{l}{Arousal(\%)} & \multicolumn{1}{l}{Valence(\%)} \\
  \midrule
  3D  & $85.49$±$0.05$ & $86.70$±$0.05$ \\
  3D+3D & $89.44$±$0.04$ & $90.59$±$0.04$ \\
  3D+1D & $88.59$±$0.04$ & $90.01$±$0.04$ \\
  3D+3D+1D & $90.56$±$0.04$ & $90.71$±$0.04$ \\
  \bottomrule
  \end{tabular}
  \end{table}
  
  The experiment results are shown in table 7, indicating that the best arrangement is two 3D with one 1D convolution layers, which achieved about 5\% higher accuracy compared to the model with one 3D convolution layer. The results show that two 3D with one 1D convolution layers can extract spatial, temporal, and functional features very well for emotion classification tasks.
  
  \subsubsection{Spatial mapping on SEED dataset}
  \
  \newline
  \indent In order to show the effectiveness of brain mapping algorithm on other dataset, different levels of spatial mapping are conducted on SEED dataset \cite{article}.
  
  The SEED dataset is similar to DEAP dataset but with more EEG channels (62) and no multi-modal data. The data in SEED dataset is transformed to be similar to data on DEAP, which includes the data shape of (45 $\times$ 45 $\times$ 62 $\times$ (128 $\times$ 63)) representing (participants $\times$ videos $\times$ channels $\times$ frames), electrodes relative positions, and keeping only negative and positive data labels.
  
  The results are shown in Table 8. In the table, 2D standard represents 2D EEG series with 62 channels, shrink-32 represents 3D mapping of 32 channels,and expand-146 represents 3D mapping of 146 EEG channels. Extra channels in expand-146 are the copies of original channels, which is filled randomly near the brain center. From the Table, the 3D spatial mapping achieves higher results than 2D EEG image series with sufficient EEG channels. This also explains the seeming ineffectiveness of spatial mapping on DEAP dataset: the lack of EEG channels (32 EEG channels on DEAP dataset).
  
  \begin{table}[htb]
  \centering
  \caption{Spatial mapping on SEED dataset}
  \begin{tabular}{cc}
  \toprule
  SEED Model name & Two category classification(\%) \\
  \midrule
  2D standard & $91.48$±$0.02$   \\
  3D shrink-32   & $88.88$±$0.04$  \\
  3D standard-62 & $97.77$±$0.01$ \\
  3D extend-146 & $99.63$±$0.01$  \\
  \bottomrule
  \end{tabular}
  \end{table}

  \section{Conclusion}
  The proposed model is proposed to improve accuracy of emotion recognition while avoiding over-fitting, which is composed of three parts: the sigmoid baseline filtering for pre-processing, the spatial and functional brain mapping of CNS and PNS signals for integrated feature representation, and compact 4D-CNN for feature extraction and emotion recognition. By combining baseline with the spatial, temporal, and functional information of various physiological signals, our proposed model can automatically extract features to accomplish the goal of emotion recognition. In addition, it is also discussed that the effectiveness of spatial and functional brain mapping, different combination of convolution layers, and potential of brain mapping on other dataset. Specifically, the accuracy of emotion classification on DEAP dataset is 92.31\% for valence and 92.76\% for arousal. Abundant experiments show that the proposed model can obtain top performance for emotion recognition.
  
  
  \newpage
  {\small
  \bibliographystyle{ACM-Reference-Format}
  \bibliography{sample-sigconf.bib}
  }
\end{sloppypar}
  \end{document}